\documentclass[12pt,preprint]{aastex}

\newcounter{address}
\newcommand{\latin}[1]{{#1}}
\newcommand{\unit}[1]{\mathrm{#1}}
\newcommand{\di}{\mathrm{d}}
\newcommand{\ie}{\latin{i.e.}}
\newcommand{\eg}{\latin{e.g.}}
\newcommand{\wpp}{w_{\mathrm{p}}}
\newcommand{\rp}{r_{\mathrm{p}}}
\newcommand{\rrf}{r_\mathrm{f}}
\newcommand{\ttdyn}{t_\mathrm{dyn}}
\newcommand{\nnLRG}{n_\mathrm{LRG}}
\newcommand{\GammaLRG}{\Gamma_\mathrm{LRG}}
\newcommand{\NNf}{N_\mathrm{f}}
\newcommand{\Gpc}{\unit{Gpc}}
\newcommand{\Mpc}{\unit{Mpc}}
\newcommand{\kpc}{\unit{kpc}}
\newcommand{\Gyr}{\unit{Gyr}}
\newcommand{\Myr}{\unit{Myr}}
\newcommand{\km}{\unit{km}}
\newcommand{\s}{\unit{s}}
\newcommand{\arcs}{\unit{arcsec}}

\newcommand{\hmpc}{h^{-1}\mathrm{Mpc}}
\newcommand{\hkpc}{h^{-1}\mathrm{Kpc}}
\newcommand{\hMsun}{h^{-1}M_{\odot}}
\newcommand{\Omegam}{\Omega_{m}}
\newcommand{\Omegab}{\Omega_{b}}
\newcommand{\Omegal}{\Omega_{\Lambda}}
\newcommand{\Mmin}{M_\mathrm{min}}

\begin{document}

\title{VERY SMALL-SCALE CLUSTERING AND MERGER RATE OF LUMINOUS RED GALAXIES}
\author{
	Morad~Masjedi\altaffilmark{\ref{NYU},\ref{email}},
	David~W.~Hogg\altaffilmark{\ref{NYU}},
	Richard~J.~Cool\altaffilmark{\ref{Steward}},
	Daniel~J.~Eisenstein\altaffilmark{\ref{Steward}},
	Michael~R.~Blanton\altaffilmark{\ref{NYU}},
	Idit~Zehavi\altaffilmark{\ref{Steward},\ref{CWRU}},
	Andreas~A.~Berlind\altaffilmark{\ref{NYU}},
	Eric~F.~Bell\altaffilmark{\ref{MPI}},
	Donald~P.~Schneider\altaffilmark{\ref{PSU}},
	Michael~S.~Warren\altaffilmark{\ref{LANL}},
	Jon~Brinkmann\altaffilmark{\ref{APO}}
}

\setcounter{address}{1}
\altaffiltext{\theaddress}{\stepcounter{address}\label{NYU} Center for
Cosmology and Particle Physics, Department of Physics, New York
University, 4 Washington Pl, New York, NY 10003}
\altaffiltext{\theaddress}{\stepcounter{address}\label{email} To whom
correspondence should be addressed:
\texttt{morad.masjedi@physics.nyu.edu}}
\altaffiltext{\theaddress}{\stepcounter{address}\label{Steward}
Steward Observatory, 933 N Cherry Ave, Tucson, AZ 85721}
\altaffiltext{\theaddress}{\stepcounter{address}\label{CWRU}
Department of Astronomy, Case Western Reserve University, Cleveland, OH 44106}
\altaffiltext{\theaddress}{\stepcounter{address}\label{MPI}
Max-Planck-Institut f\"ur Astronomie, K\"onigstuhl 17, D-69117
Heidelberg, Germany}
\altaffiltext{\theaddress}{\stepcounter{address}\label{PSU} Department
of Astronomy and Astrophysics, Pennsylvania State University, 525
Davey Laboratory, University Park, PA 16802}
\altaffiltext{\theaddress}{\stepcounter{address}\label{LANL}
Theoretical Division, Los Alamos National Laboratory, Los Alamos, NM
87545} \altaffiltext{\theaddress}{\stepcounter{address}\label{APO}
Apache Point Observatory, P.O. Box 59, Sunspot, NM 88349}

\begin{abstract}
We present the small-scale ($0.01<r<8\,h^{-1}~\Mpc$) projected
correlation function $\wpp(\rp)$ and real space correlation function
$\xi(r)$ of $24520$ luminous early-type galaxies from the Sloan
Digital Sky Survey Luminous Red Galaxy (LRG) sample (
$0.16<z<0.36$). ``Fiber collision'' incompleteness of the SDSS
spectroscopic sample at scales smaller than $55~\arcs$ prevents
measurements of the correlation function for LRGs on scales smaller
than $\sim 0.3~\Mpc$ by the usual methods. In this work, we
cross-correlate the spectroscopic sample with the imaging sample, with
a weighting scheme to account for the collisions, extensively tested
against mock catalogs. We correct for photometric biases in the SDSS
imaging of close galaxy pairs. We find that the correlation function
$\xi(r)$ is surprisingly close to a $r^{-2}$ power law over more than
4 orders of magnitude in separation $r$. This result is too steep at
small scales to be explained in current versions of the halo model for
galaxy clustering. We infer an LRG-LRG merger rate of $\lesssim
0.6~\Gyr^{-1}\Gpc^{-3}$ for this sample. This result suggests that the
LRG-LRG mergers are not the main mode of mass growth for LRGs at
$z<0.36$.

\end{abstract}


\section{INTRODUCTION}

Galaxy clustering is a tool for the study of a diverse set of
phenomena at different scales. On large scales ($\sim 100~\Mpc$) the
density perturbations are small enough to be described in a linear
framework, allowing analysis of cosmological models in detail and
constraints on cosmological parameters \citep{tegmark03b}. In
addition, recent studies have used extremely large-scale galaxy
clustering to test fundamental cosmological hypotheses, such as
homogeneity of the Universe \citep{hogg05a} and flatness of the
universe \citep{eisenstein05b}.

On intermediate scales ($0.3$ to $30 ~\Mpc$), galaxy clustering probes
the relation of galaxies to dark matter through the biased clustering
of dark matter halos. On these scales the two-point correlation
function is found to be very close to a power law. Recent work has
shown that the observed small deviations of the correlation function
from a power-law on these scales can be interpreted in the framework
of the Halo Occupation Distribution \citep[HOD;][]{peacock00a,
seljak00a, scoccimarro01a, berlind02a} as a natural transition between
galaxy pairs within a single virialized halo and galaxy pairs in
separate halos \citep[\eg, ][]{zehavi04a,zehavi05b}.

On sufficiently small scales within a halo ($< 1~\Mpc$), galaxy
clustering will probe more than just the properties of the dark
matter. On these scales one expects a variety of more complex
processes to modify the galaxy clustering and thus presumably give
rise to features in the correlation function $\xi(r)$. These processes
include dynamical friction, tidal interactions, stellar feedback and
other dissipative processes. These processes will force close galaxy
pairs to merge over time-scales of order of the dynamical time ($\sim
10^8~\mathrm{yr}$) \citep{toomre77a, larson78a, vandokkum05a,
bell05a}. Thus measurements of the correlation function $\xi(r)$ on
small scales can be translated to a number density of near-future
merger events. Some understanding of the merger process can be used to
turn this density into a galaxy merger rate or at least a constraint
thereon, providing an empirical measure of the importance of mergers
as a fundamental mode of stellar mass addition to galaxies.

In addition, numerical simulations within the cold dark matter
paradigm find that self-bound substructures within dark matter halos
(subhalos) merge to form larger systems. It is tempting to associate
these subhalos with galaxies in groups and clusters \citep{klypin99a,
kravtsov99a, colin99a, moore98a}. Comparison between galaxy clustering
and subhalo clustering on very small scales can lead to a better
understanding of the connection between galaxies and dark matter
subhalos \citep{kravtsov04b,zentner05a}.

There are no measurements of the real-space correlation function on
scales smaller than $ 100~\kpc$. This is mainly due to the fact that
on small angular separations the selection function of typical surveys
becomes too complex and the pair counts are limited by shot
noise. \cite{gott79a} and \cite{maller05a} measured the angular
correlation function for angular separations that, for the median
redshift of their sample translates to $10<r<100~ \kpc$. They both
find that the correlation function $\xi(r)$ power law extends down to
these small scales. The drawback of these studies is that they measure
only angular, not proper correlations, and they are difficult to
interpret without precise knowledge of the radial selection function
and density field.

The Sloan Digital Sky Survey \citep[SDSS;][]{york00a} has provided the
largest spectroscopic sample ever of these massive galaxies; its
special Luminous Red Galaxy (LRG) target selection
\cite{eisenstein01a} uses color to pre-select luminous early-type
galaxies in a volume much larger than that of the SDSS Main sample.
Because of their high luminosities, their association with massive
halos, and their spectral uniformity, LRGs are extremely useful probes
of large-scale structure.  In addition, however, their morphological
simplicity makes it straightforward to study their clustering at very
small scales; Their lack of cold gas (and, indeed, the lack of equally
bright blue galaxies) means that close passages and mergers do not
trigger significant star formation, change the colors significantly,
or change the morphology significantly, and therefore do not
``remove'' systems of interest at small scales from the parent sample.
In this paper we capitalize on these properties of the LRGs---and the
enormous volume of the SDSS LRG sample---to measure the clustering to
extremely small scales ($\sim 10~\kpc$) and to constrain the LRG--LRG
merger rate.
 
Throughout this paper all distances are comoving, calculated for a
cosmological world model with
$(\Omega_\mathrm{m},\Omega_\Lambda)=(0.3,0.7)$ and Hubble constant
parameterized by $H_0\equiv 100~h\,\km\,\s^{-1}\,\Mpc^{-1}$. At the
mean redshift of the sample ($z\sim 0.3$), the fiber collision scale,
$55~\arcs$, corresponds to $\sim 200~\kpc$ and the SDSS median PSF
width, $1.4~\arcs$, translates to $\sim 6~ \kpc$.

\section{DATA}

The SDSS \citep{stoughton02a,abazajian03a,abazajian04a} is conducting
an imaging survey of $\sim 10^4$ square degrees in 5 bandpasses: $u$,
$g$, $r$, $i$, and $z$ \citep{fukugita96a,gunn98a,
gunn05a}. Photometric monitoring \citep{hogg01a}, image processing
\citep{lupton01a,stoughton02a,pier03a}, and good photometric
calibration \citep{smith02a, ivezic04a} allow one to select galaxies
\citep{strauss02a,eisenstein01a}, quasars \citep{richards02a}, and
stars for spectroscopic observations with the twin fiber-fed
double-spectrographs.

Targets are assigned to spectroscopic fiber plug plates with a tiling
algorithm that ensures nearly complete samples \citep{blanton03a}. The
angular completeness is characterized carefully for each unique region
of overlapping spectroscopic plates (``sector'') on the sky. An
operational constraint of SDSS spectrographs is that the physical size
of the fiber coupling forces the angular separation of the targets to
be larger than $55~\arcs$. This ``fiber collision`` constraint is partly
reduced by having roughly one third of the sky covered by overlapping
plates, but it still results in $\sim 7\%$ of targeted galaxies not
having measured redshifts.

We focus here on the Luminous Red Galaxy spectroscopic sample
\citep{eisenstein01a}. This sample is constructed from color-magnitude
cuts in $g$, $r$, and $i$ to select galaxies that are likely to be
luminous early-type galaxies at redshifts between 0.15 and 0.5. The
selection is highly efficient and the redshift success rate is
excellent. The sample is constructed to be close to volume-limited up
to $z=0.36$, with a dropoff in density toward $z=0.5$.

This study uses a sample drawn from NYU LSS {\tt sample14}
\citep{blanton05a} and covers 3,836 square degrees containing 55,000
LRGs between redshift of 0.16 and 0.47. The subsample of LRGs used in
this paper has luminosity and redshift ranges of $- 23.2<M_g<-21.2$
and $0.16<z<0.36$, respectively. The absolute magnitudes include
Galactic extinction corrections \citep{schlegel98a}, $k$-corrections
and passive evolution corrections to redshift $z=0.3$. This subsample,
which is chosen to maximize our use of the volume-limited portion of
the LRG spectroscopic sample, is identical to the first subsample
(29298 galaxies) used in \cite{zehavi05a}, with the
difference that in this paper we limit ourself only to the North
Galactic Cap (24520 galaxies). This is due to the slight difference in
some aspects of tiling and photometric calibrations between the North
and South Galactic Cap areas. The details of the radial and angular
selection functions are described elsewhere \citep{zehavi05a}.

We create large catalogs of randomly distributed points based on these
angular and radial models. These catalogs match the redshift
distribution of the LRGs and are isotropic within the survey region.
These catalogs allow us to check the survey completeness of any given
volume and provide a homogeneous baseline (\eg, expected numbers) for
the tests that follow.

\section{METHOD \& RESULTS}
\subsection{Projected Correlation Function}
To calculate the real-space correlation function on intermediate
scales, one estimates the correlation function on a two-dimensional
grid of pair separations parallel ($\pi$) and perpendicular ($\rp$) to
the line of sight, termed $\xi(\rp,\pi)$. This can be turned into
projected correlation function

\begin{equation}\label{eq:wp}
 \wpp(\rp)=2\int_0^{\pi_{max}}\di\pi\,\xi(\rp,\pi).
\end{equation}
in which $\pi_{max}$ is set to a value sufficiently large to include
most correlated pairs and give stable results (\ie, independent of the
choice of $\pi_{max}$ with in the error-bars). Using this method,
\cite{zehavi05a} calculate $\wpp(\rp)$ for the LRG sample in the range
$0.3~\Mpc \lesssim \rp \lesssim 30~\Mpc$. The lower limit in the
\cite{zehavi05a} analysis was set to eliminate the incompleteness effects
that are introduced by the $55~\arcs$ fiber collision radius.

We overcome the fiber collision problem by cross correlating the
spectroscopic sample with the full sample of LRG targets in the SDSS
imaging, whether or not they have observed redshifts. To do this we
use the \cite{zehavi05a} subsample of LRGs (as explained above) as our
spectroscopic sample, and \emph{all} LRG targets as our imaging
sample. For each LRG from the spectroscopic sample, we treat the
nearby imaging LRGs (whether they have spectrum or not) as if they are
at the same redshift as the spectroscopically observed LRG. This allows us to
calculate a $g$-band absolute magnitude ($M_g$) for them in the same
manner as the LRGs with spectra and choose the ones that make it
inside the sample limits (\eg, $- 23.2<M_g<-21.2$). This turns our
cross-correlation into an auto-correlation. In addition, assuming the
LRG redshift we can bin the pairs according to their comoving
projected separation.

We statistically remove the interlopers, \ie, galaxies not at the same
redshift but treated as such by the algorithm, by making random
spectroscopic samples with the same redshift distribution as LRG
spectroscopic sample and cross-correlating this sample with the
imaging LRG sample. We subtract the scaled random interlopers from our
data-data correlation.

This method makes use of the fact that essentially all the support for
the projected correlation function $\wpp$ integral (\ref{eq:wp}) arises
from line-of-sight separations $\pi$ that are much smaller than the
distance to the given galaxy. The downside to this method is the loss
of signal-to-noise due to interlopers, which becomes irrelevant at
small-scales, where the correlation function becomes much larger than
unity.

Using this method we calculate $\wpp(\rp)$; schematically:
\begin{equation}
n\,\wpp(\rp)=\frac{\mathrm{D_sD_i}}{
\mathrm{D_sR_i}}-\frac{\mathrm{R_sD_i}}{ \mathrm{R_sR_i}} \quad ,
\end{equation}
where $n$ is the average comoving density of the spectroscopic LRG
sample used here, $\mathrm{D_s}$ and $\mathrm{D_i}$ are the
spectroscopic and imaging data samples, and $\mathrm{R_s}$ and
$\mathrm{R_i}$ are the random spectroscopic and random imaging
samples. In detail, the factors are:

\begin{equation}
\mathrm{D_sD_i}=\frac{\displaystyle\sum_{j \in D_sD_i
\mathrm{pairs}}p_j}{\displaystyle \sum_{j \in {D_s}}p_j} \quad ,
\end{equation}
where $p_j$ is the weight given to each spectroscopic LRG to account
for the fiber collisions. We calculate this weight by running a
friends-of-friends grouping algorithm with a $55~\arcs$ linking length
provided within {\tt sample14} package. Within each ``collision
group'' made by the friends-of friends we find the number of objects
that with spectroscopic redshifts and divide by the total number. The
inverse of this ratio is the weighting $p_j$ assigned to each
spectroscopic LRG. This procedure emulates the SDSS tiling algorithm
\citep{blanton03a}.

\begin{equation}\label{eq:secterm}
\mathrm{D_sR_i}=\frac{\displaystyle \sum_{j \in D_sR_i
\mathrm{pairs}}p_j}{\displaystyle \sum_{j \in {D_s}}p_j
\left(\frac{\di \Omega}{\di A}\right)_j \frac{\di N}{\di \Omega}} \quad ,
\end{equation}
where $\left(\frac{\di \Omega}{\di A}\right)_j$ is the inverse square
of the comoving distance to spectroscopic galaxy $j$ and $\frac{\di
N}{\di \Omega}$ is the number density of the random imaging catalog
per solid angle. The average of multiplication of these two terms
gives the average number of random imaging objects per unit comoving
area around each spectroscopic galaxy.

\begin{equation}
\mathrm{R_sD_i}=\frac{\displaystyle\sum_{j \in R_sD_i
\mathrm{pairs}}f_j}{\displaystyle \sum_{j \in \mathrm{R_s}}f_j}
\quad ,
\end{equation}
where $f_j$ is the weight given to spectroscopic galaxy $j$ which
accounts for the incompleteness of the spectroscopic survey in that
region of the sky \emph{not} due to fiber collision but due to all the
other selection effects in survey.  The {\tt sample14} package
provides the angular geometry of the spectroscopic survey expressed in
terms of spherical polygons.  The geometry is complicated: the
spectroscopic plates are circular and overlap, while the imaging is in
long strips on the sky, and there are some overlap regions of certain
plates that may not have been yet observed. The resulting spherical
polygons track all these effects and characterize the geometry in
terms of ``sectors'', each being a unique region of overlapping
spectroscopic plates. In each sector, we count the number of possible
targets (LRG, MAIN, and quasar), excluding those missed because of
fiber collisions, and the number of these whose redshifts were
determined. We weight the random spectroscopic LRGs by the inverse of
the ratio of these numbers ($f_j$). In truth, the priority of all
targets are not equal, such that LRGs will always lose to quasar
candidates, but the LRG priority is equal to the dominant MAIN
targets. Only about $12\%$ of the fibers are assigned to quasars,
hence quasar LRG collisions are rare and this priority bias could be
ignored. 

\begin{equation}
\mathrm{R_sR_i}=\frac{\displaystyle\sum_{j \in R_sR_i
\mathrm{pairs}}f_i}{\displaystyle \sum_{j \in {R_s}}f_j
\left(\frac{\di \Omega}{\di A}\right)_j \frac{\di N}{\di \Omega}}
\quad ,
\end{equation}
is similar to equation \ref{eq:secterm} but for the
random Spectroscopic and random imaging pairs.

\begin{figure}
\begin{center}
\includegraphics[width=1\textwidth]{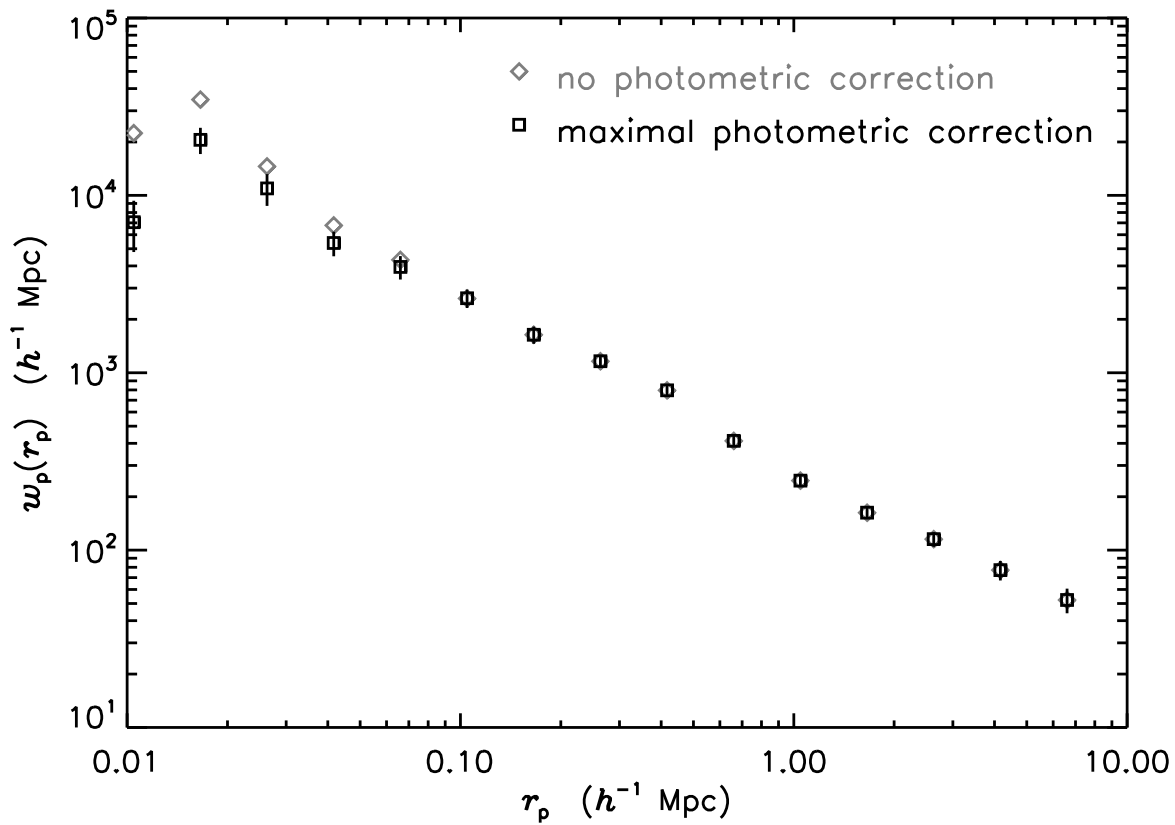}
\caption{Projected correlation function $\wpp(\rp)$ for the
 LRG sample ($-23.2<M_g<-21.2$ and $0.16<z<0.36$)
 calculated as described in the text. The gray diamonds show the
 measured projected correlation function before correction for the
 photometric bias in the close galaxy pairs. The error-bars are from
 the jackknife error covariance matrix. }
\label{fig:w_rp}
\end{center}
\end{figure}

Figure ~\ref{fig:w_rp} shows the projected correlation function
measured in this manner (data in Table \ref{tab:data}). The error-bars
are estimated using jackknife resampling covariance matrix with 50
subsamples.These results agree well with \cite{zehavi05a} on their
overlap range ($0.3$ to $8\,h^{-1}~\Mpc$) and on smaller scales
($0.01$ to $0.3\,h^{-1}~\Mpc$) is very close to an extension of the
best-fit power-law from \cite{zehavi05a}.

\subsection{Test on Mock LRG Catalogs}

To test our method, we use an N-body simulation of a $\Lambda$CDM
cosmological model, with $\Omegam=0.3$, $\Omegal=0.7$, $\Omegab=0.04$,
$h\equiv H_0/(100~\mathrm{km~s}^{-1}~\mathrm{Mpc}^{-1})=0.7$,
$n_s=1.0$, and $\sigma_8=0.9$.  This model is in good agreement with a
wide variety of cosmological observations (see, e.g.,
\citealt{spergel03a,tegmark04a, abazajian05b}). Initial conditions
were set up using the transfer function calculated for this
cosmological model by CMBFAST \citep{seljakzal96}.  The simulation was
run at Los Alamos National Laboratory (LANL) using the Hashed-Oct-Tree
(HOT) code \citep{warren91a}; the simulation followed the evolution of
$1024^3$ dark matter particles, each of mass $3.51\times
10^{10}\hMsun$, in a comoving box of size $768\hmpc$.  The
gravitational force softening is $\epsilon_{\rm grav}=12\hkpc$
(Plummer equivalent). We identify halos in the dark matter particle
distributions using a friends-of-friends algorithm with a linking
length equal to $0.2$ times the mean inter-particle separation.  We
populate these halos with galaxies using a simple model for the Halo
Occupation Distribution (HOD).  Every halo with a mass $M$ greater
than a minimum mass $\Mmin$ is assigned a central galaxy that is placed at
the halo center of mass and is given the mean halo velocity.  A number
of satellite galaxies is then drawn from a Poisson distribution with
mean $((M-\Mmin)/M_1)^\alpha$, for $M\geq\Mmin$.  These satellite
galaxies are assigned the positions and velocities of randomly
selected dark matter particles within the halo.  We select the
parameter values $\Mmin=4.5\times10^{13}\hMsun$,
$M_1=3.5\times10^{15}\hMsun$, and $\alpha=1$ that yield a mock galaxy
population with the observed space density of LRGs and approximately
the correct galaxy-galaxy correlation function.

In order to carve the SDSS LRG sample geometry out of our mock cube,
we create a new cube with 27 times larger volume by tiling the mock
cube $3\times3\times3$.  Since the N-body simulation used to construct
the mock was run with periodic boundary conditions, we can tile the
cube without having density discontinuities at the boundaries.  We set
the center of this tiled cube to be the origin and put galaxies into
redshift space using the line-of-sight component of their peculiar
velocities.  We then compute RA, Dec, and redshift coordinates for
every mock galaxy in the tiled cube.  Finally, we only keep galaxies
whose coordinates would place them within the LRG sample geometry.
The final step in creating a mock LRG catalog is to incorporate the
fiber collision constraint.  In the SDSS, LRG fibers collide both with
each other, and with other galaxies that are mostly uncorrelated
foreground galaxies.  We create a foreground screen of mock galaxies
on the sky by populating the same N-body simulation with a different
HOD that yields an angular correlation function equal to the mean for
all SDSS galaxies.  We then allow all galaxies to collide with each
other and keep track of collided mock LRG galaxies. This approach
produces results that are very close to the SDSS tiling algorithm
\citep{blanton03a}.

\begin{figure}
\begin{center}
\includegraphics[width=1\textwidth]{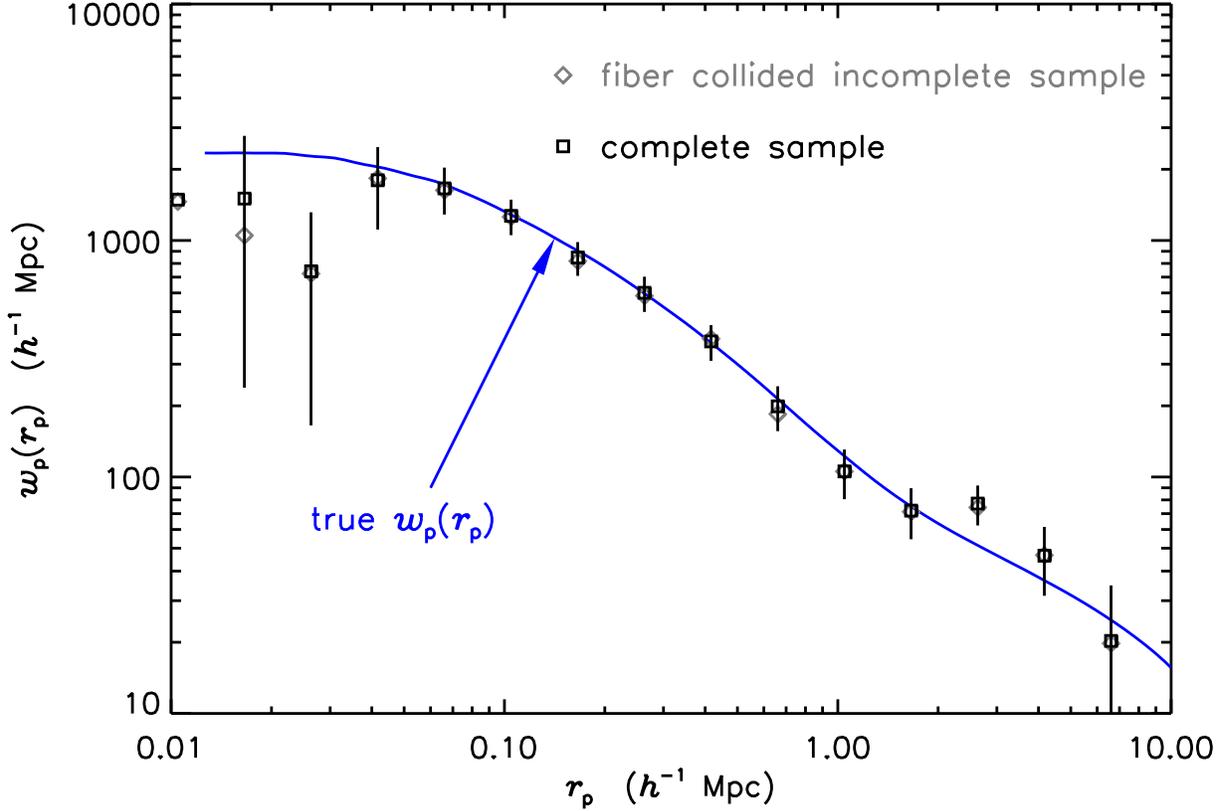}
\caption{Projected correlation function $\wpp(\rp)$
measured for the mock cube measured directly in the simulation (solid
line), and by the method described in the text after applying SDSS
geometry cuts(black squares). The gray diamonds show the measured
$\wpp(\rp)$ for the mock after applying the SDSS geometry cuts and the
the fiber collision incompleteness. It is worth noting that the
geometry cuts select about a third of all galaxies in the cube
therefore on very small scales the data points are dominated by the
shot noise.}

\label{fig:mock}
\end{center}
\end{figure}

We use this mock to test two different aspect of our correlation
estimator: the cross-correlation method and the fiber collision
weighting scheme. First we use all the mock galaxies as both our
spectroscopic and imaging samples (no galaxies eliminated by fiber
collisions). Next we drop the galaxies that would not get a redshift
in SDSS due to fiber collision from our spectroscopic sample and
repeat the test. Figure \ref{fig:mock} shows the measured projected
correlation function for both cases in comparison to the real
projected correlation function for the full cube of the mock. It is
worth noting that on very small scales the shot noise dominates the
galaxy pair counts and limits the comparison.

\subsection{Photometry Test}

One important issue with all clustering measurements on small scales
is possible photometric biases when measuring close pairs. This may
emerge when the outskirts of a pair of galaxies overlap on these
scales. This can lead to biased flux measurements for galaxies which
will affect the completeness of the samples. To estimate the magnitude
of this effect for SDSS, we ran a simulation in which we create fake
images of pairs of galaxies with separations ranging from $2$ to
$35~\arcs$. We studied two different cases, one for galaxy pairs
consisting of two identical galaxies and another with galaxies of
different luminosities. The basic information regarding these galaxies
is summarized in Table \ref{tab:sim}. These galaxies represent
passively evolving LRG galaxies observed at a redshift of $z=0.3$ with
de Vaucouleurs profiles ($n=4$ S$\mathrm{\acute{e}}$rsic profiles).

We place one such galaxy pair onto RUN $2662$ of SDSS imaging. This
RUN has a typical SDSS seeing of about $1~\arcs$. Each galaxy image is
convolved with the local seeing of the field. The flat fielding
vectors and gain corrections applied to the SDSS imaging in reverse to
each constructed image. After inserting known bad pixels into the mock
galaxy images; these images are added to raw SDSS images. These new
images are then processed using the standard SDSS pipeline, PHOTO, to
determine the effect of proximity of galaxies on their measured
properties.

\begin{figure}
\begin{center}
\includegraphics[width=1\textwidth]{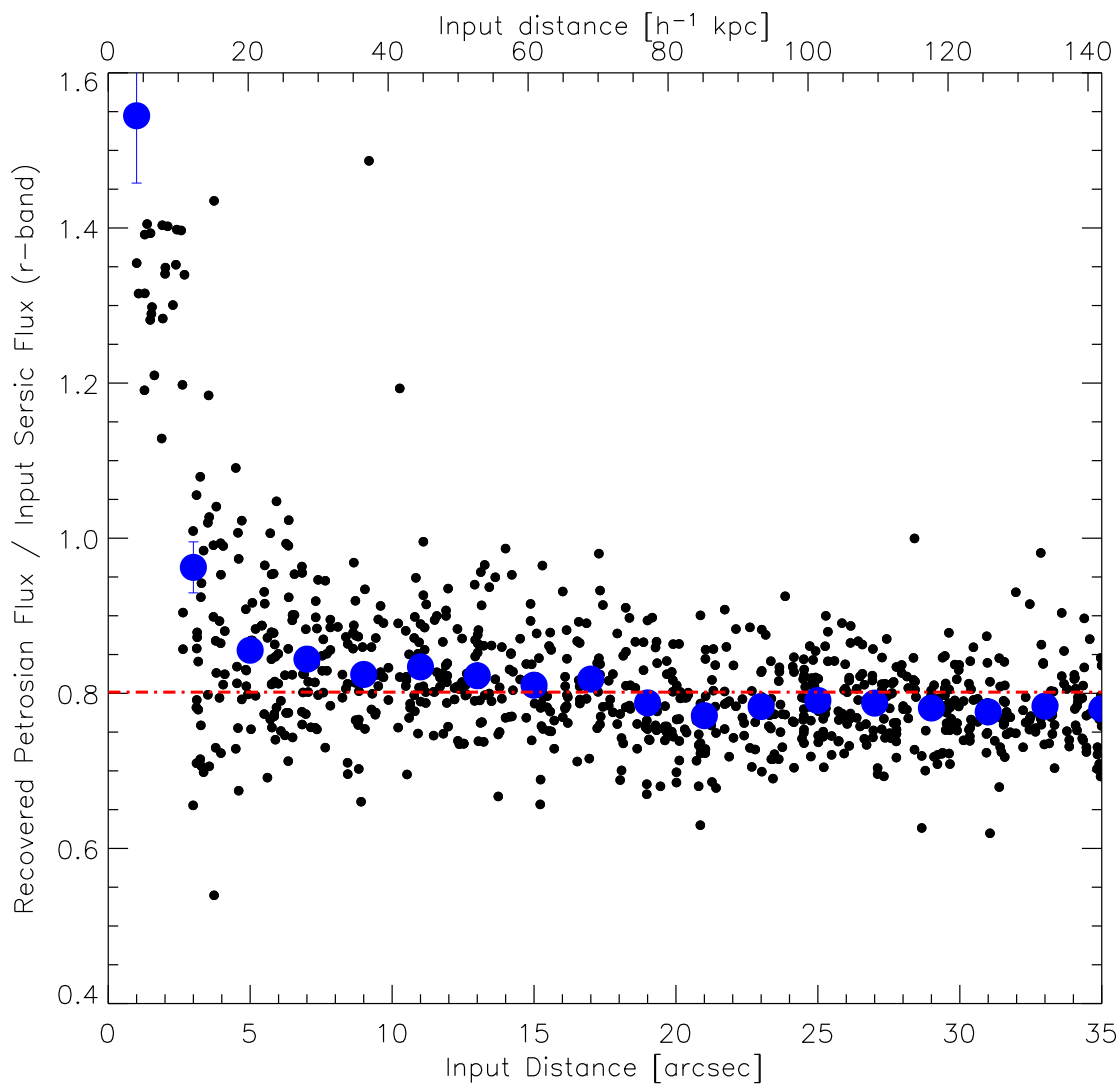}
\caption{Recovered Petrosian flux to input S$\mathrm{\acute{e}}$rsic
flux as a function of the separation of the two galaxies in the pair,
the blue dots show the three sigma out-layer rejected average of the
recovered flux for different separations. It can be seen that the
pipeline completely fails for galaxies closer than $3~\arcs$ and on
average there is an excess in the recovered flux of galaxies separated
by less than $20~\arcs$. }
\label{fig:richard}
\end{center}
\end{figure}

We found that for the case of two identical galaxies, for separations
below $3~\arcs$, the two galaxies are not well deblended, leading to
the detection of a single galaxy with combined flux of the individual
galaxies. At separations larger than $20~\arcs$, the Petrosian flux
measures $79.5$ percent of the input S$\mathrm{\acute{e}}$rsic flux
(as expected) and the ratio of the recovered to input flux is
independent of the separation. In other words, in the absence of a
close neighbor for a $n=4$ S$\mathrm{\acute{e}}$rsic galaxy, the
Petrosian flux only measures about $80$ percent of a galaxy's
light. For intermediate separations, ($ 5 < s < 20~\arcs$), the
fraction of the recovered flux to input flux increases to $83$
percent.  This increase is likely due to a double counting of the low
level diffuse emission from the two galaxies which is being poorly
deblended between the two objects. Figure \ref{fig:richard} shows this
result.

Using this result we correct for the completeness of our sample on
small scales. We convert angular separation to projected separation
using the median redshift of the sample $z=0.3$ (the same redshift is
used for the test). In addition we know from \cite{eisenstein01a} that
for each $0.1$ magnitude change of the faint limit of our sample the
number of galaxies in the samples is increased by roughly 30
percent. For each projected distance separation $\rp$ bin in the
projected correlation function $\wpp(\rp)$ we convert the excess flux
count for that separation to an excess of galaxy counts in the
sample. The square of this quantity will be the excess in galaxy pairs
allocated to that separation bin. we then correct our projected
correlation function by this factor for each given separation (Figure
\ref{fig:w_rp}, Table \ref{tab:data}). The result for the second case
with galaxies of different magnitudes is close enough to the first
case that we use only the first case to correct our sample.

\subsection{Real-Space Correlation Function}

The projected correlation function $\wpp(\rp)$ can be ``deprojected''
to get $\xi(r)$ by

\begin{equation}
\xi(r)=-\frac{1}{\pi}\int^{\infty}_r d\rp
\frac{d\wpp(\rp)}{d\rp}(\rp^2-r^2)^{-1/2}
\end{equation}
\citep[\eg,][]{davis83a}. We calculate this integral analytically
by linear interpolation between the binned $\wpp(\rp)$ values,
following \cite{Saunders92a}. This estimate is only accurate to
a few percent, due to limitations of the interpolation.

Figure ~\ref{fig:xi} shows the real-space correlation function,
obtained in this fashion, combined with the real-space correlation
function $\xi(r)$ on intermediate-scales from \cite{zehavi05a} and
redshift-space correlation function $\xi(s)$ on large-scales from
\cite{eisenstein05b} for the LRG sample. Also shown are the power-law
$\xi(r)=[r/(10\,h^{-1}~Mpc)]^{-2.0}$ and the ``1-halo term'' of the
correlation function (which only counts pairs of galaxies within the
same dark matter halo) calculated for the HOD parameters given by
\cite{zehavi05b} for the $M_r<-22$ SDSS MAIN sample which is close to
the LRG sample.

Figure ~\ref{fig:rsqxi} shows the real space correlation function
divided by a $r^{-2}$ power-law to accentuate the deviations from a
power-law. The dip at $1~\Mpc$ is described and quantified by the halo
model as the transition from 2-halo to 1-halo term
\citep{zehavi04a}. The upturn at $0.03~\Mpc$ could be real but is not
highly significant. Finally the drop of the innermost point at
$0.01~\Mpc$ is most probably due to deblending issues.

\subsection{Merger Rate}

If we interpret the LRG correlation function $\xi(r)$, measured at
small scales as a quasi--steady-state inflow leading to the mergers of
pairs of LRGs, we can straightforwardly turn the measured $\xi(r)$
into a merger rate.

We assume that there is a length scale $\rrf$ inside of which dynamical
friction is so effective that pairs at this separation merge in a
dynamical time $\ttdyn$, where the dynamical time is
\begin{equation}
\ttdyn \approx \frac{2\,\pi\,\rrf}{v_{circ}} \quad ,
\end{equation}
and $v_{circ}$ is the circular velocity of the orbit, which is roughly
$1.5$ times the velocity dispersion $\sigma_v$; the exact value of the
numerical factor depends on the velocity ellipsoid. For typical
velocity dispersions, $\sigma_v\sim 200~\km\,\s^{-1}$, and merger
length scales,$\rrf \sim 10~\kpc$, we find a dynamical time $\ttdyn\sim
200~\Myr $, which is in agreement with time-scales derived by
\cite{bell05a} from merger simulations.

If the number density of LRGs is $\nnLRG$, the average number $\NNf$
within distance $\rrf$ of any ``target'' LRG is
\begin{equation}
\NNf \approx 4\,\pi\,\nnLRG \,\int_0^{\rrf}\,r^2\,\xi(r)\,\di r \quad .
\end{equation}
and the LRG merger rate $\GammaLRG$ is, by assumption
\begin{equation}
\GammaLRG = \frac{\NNf}{\ttdyn}
   \approx 3\,\rrf^2\,\xi(\rrf) \,\nnLRG \, \sigma_v
   \quad .
\end{equation}

In this derivation we have used the fact that the $\xi(r)\sim r^{-2}$
(Figure ~\ref{fig:rsqxi}). This merger rate can be written as
\begin{equation}
\GammaLRG \approx \frac{1}{160~\Gyr}
   \,\left[\frac{\rrf^2\,\xi(\rrf)}{100~\Mpc^{2}}\right]
   \,\left[\frac{\sigma_v}{200~\km\,\s^{-1}}\right]
   \,\left[\frac{\nnLRG}{10^{-4}~\Mpc^{-3}}\right]
   \quad ,
\end{equation}
which is equivalent to a comoving volume merger rate of:
\begin{equation}
\dot{\phi}_M\equiv \frac{\GammaLRG}{\nnLRG} \sim 0.6 \times
10^4~\Gyr^{-1}\,\mathrm{Gpc}^{-3}\quad .
\end{equation}

This merger rate does not depend on the choice of $\rrf$ which is
indeed poorly known. It is worth noting that this merger rate is a
strict upper limit. First, we assume that all close pairs will
merge, but in principle galaxies in high density clusters could
pass by each other without merging. Secondly, we are assuming that
all galaxies will merge in a dynamical time, which is the minimum
possible time for a merger to occur.

\section{DISCUSSION}

We have combined the data for $24520$ luminous red galaxies from the
SDSS spectroscopic sample with LRG targets in the SDSS imaging sample
to measure the strength of clustering for LRGs on small scales ($0.01$
to $8\,h^{-1}~\Mpc$). We deal with the fiber collision incompleteness
of the SDSS spectroscopic sample on scales smaller than $55~\arcs$ by
cross-correlating the spectroscopic and imaging samples and
statistically removing the interloping galaxies. This method is
extremely powerful on small scales at which the correlation function
becomes much larger than unity.

We find that the correlation function on these scales is very close to
an extrapolation of the correlation function power-law found on larger
scales $\xi(r)\propto r^{-2}$ \citep{zehavi05a}. This is surprising, as
one might expect the direct interactions between galaxies (\eg,
dynamical friction, galaxy merger, tidal impulses, etc.) to create
features in the correlation function.

\begin{figure}
\begin{center}
\includegraphics[width=1\textwidth]{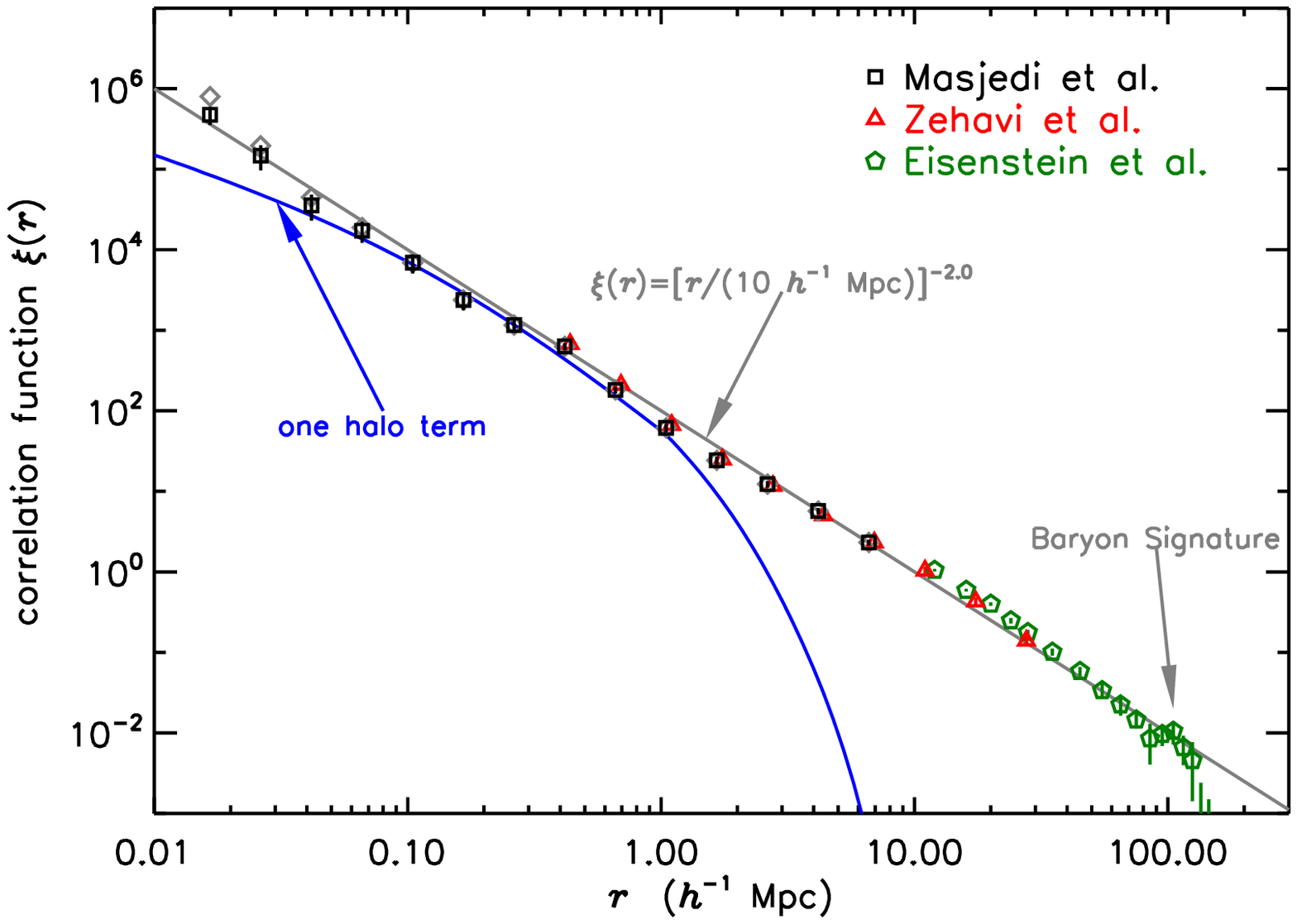}
\caption{Real-space correlation function $\xi(r)$ for the LRG sample
($-23.2<M_g<-21.2$ and $0.16<z<0.36$) calculated as described in the
text on small scales, combined with real-space correlation function on
intermediate scales from \cite{zehavi05a} and redshift-space
correlation function $\xi(s)$ on large scales from
\cite{eisenstein05b} (data points from Zehavi results are shifted by
$5\%$ in the radial direction for illustration purposes). The gray
diamonds show the result without photometric correction as in figure
\ref{fig:w_rp}. The Blue line shows the 1-halo term of the correlation
function calculated for the HOD parameters given by \cite{zehavi05b}.}
\label{fig:xi}
\end{center}
\end{figure}

\begin{figure}
\begin{center}
\includegraphics[width=1\textwidth]{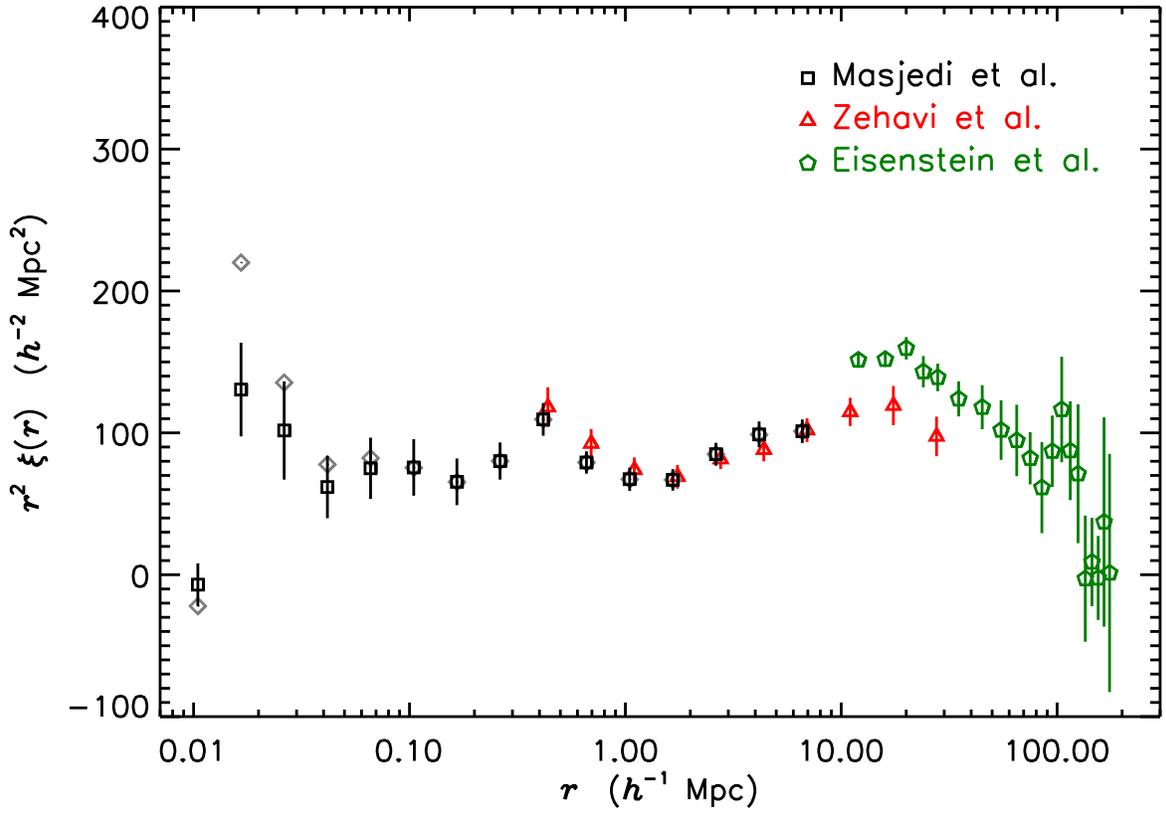}
\caption{Same as Figure \ref{fig:xi}, but $\xi(r)$ divided by a $r^{-2}$
power-law to accentuate the deviations from a power-law. Note that the
difference between \cite{zehavi05a} and \cite{eisenstein05b} is solely
due to the difference between redshift space and real-space
correlation functions.}
\label{fig:rsqxi}
\end{center}
\end{figure}

This result cannot be simply explained with the current best-fit of
the of Halo Occupation Distribution (HOD) models for galaxy clustering
\citep[\eg,][]{zehavi05a,tinker05a}. This inconsistency arises from
the fact that in current HOD models the galaxy-galaxy correlation
function on small scales simply follows the convolution of the dark
matter halo profile with itself (1-halo term). For an NFW profile
\citep{navarro97a} this relation is proportional to $r^{-1}$ toward
the core of the halo and similarly for the Moore profile
\citep{moore98b} produces a $r^{-1.5}$ relation; neither is
sufficiently steep to fit the results of this paper. Figure
~\ref{fig:xi} shows the 1-halo term of the correlation function
calculated for the HOD parameters given in \cite{zehavi05b}.  If HOD
models are modified to have a galaxy distribution much more
concentrated than the dark matter or to have dark matter halos with
density profiles much steeper than NFW toward the core of the halo,
this result could in principle be accommodated within HOD models.

We convert the correlation function on these scales to statistics of
galaxies in close dynamical pairs, and we infer an LRG-LRG merger rate
of $\lesssim 1/160~\Gyr^{-1}$ or a comoving volume merger rate,
$\dot{\phi}_M \sim 0.6 \times 10^4~\Gyr^{-1}\,\mathrm{Gpc}^{-3}$. The
fact that in large clusters, galaxies can closely pass each other
without merging, turns any merger rate inferred from close dynamical
pair statistics into an strict upper limit. This upper limit on the
LRG merger rate could in principle be violated if the merger process
triggers such strong star-formation activities in these galaxies that
they become too blue to make it to the LRG sample selection
cuts. However, this proposal is not tenable, as LRGs contain little
apparent gas or dust and essentially there are no blue galaxies with
luminosities comparable to the LRGs
\citep{blanton03d,eisenstein01a}. Therefore the LRG merger rate is in
fact very close to merger rate for all galaxies with LRG luminosities
regardless of their color.

In previous work, there are two main methods for measuring merger
rates, the most popular one is to convert the close pair statistics to
a merger rate \citep[\eg,][]{zepf89a, burkey94a, patton97a, patton00a,
carlberg00clust,bell05a}. The other method uses the asymmetry
parameter ($A$) to determine the fraction of galaxies undergoing
mergers \citep{conselice03a, conselice03b, vandokkum05a}. It is
difficult to compare these previous studies of the merger rate with
that found here because our LRG sample is more luminous and
has a significantly smaller number density than the samples uses in
those studies. The merger rate has been found to be higher for higher
redshifts and higher for fainter samples \citep{conselice03a}. For
this reason, an SDSS-size sample was required to measure the merger
rate for LRG type galaxies at low redshifts. Nevertheless our result
is, in principal, consistent with the extrapolation of the best fits to
samples at higher redshifts or fainter samples \citep{conselice05a}.

One question to address by merger-rate measurements is the importance
of major mergers in the mass build-up of massive galaxies. This could
be answered by measuring the cross-correlation of galaxies of
different mass with LRGs on small scales to obtain a mass spectrum for
objects merging into LRGs. \cite{murali02a} created a simulation for
$L^{*}$ galaxies to answer this question. They find that smooth
accretion plays the dominant role in mass build-up. Unfortunately the
resolution of the simulation ($L\sim L^{*}/4$) was too close to the
galaxies in question to be able to distinguish true smooth accretion
from mergers with objects below the mass resolution. LRGs are massive
enough that the merger of objects far below their mass could be
studied in hydro-dynamical simulations to give a detailed mass
spectrum for merger events.

\acknowledgements It is a pleasure to thank Jim Peebles, Sebastian
Pueblas, David Schlegel, and Roman Scoccimarro for valuable
discussions and software.  MM, DWH, MRB, and AB are partially
supported by NASA (NAG5-11669) and NSF (AST-0428465). This research
made use of the NASA Astrophysics Data System.

Funding for the creation and distribution of the SDSS Archive has been
provided by the Alfred P. Sloan Foundation, the Participating
Institutions, the National Aeronautics and Space Administration, the
National Science Foundation, the U.S. Department of Energy, the
Japanese Monbukagakusho, and the Max Planck Society. The SDSS Web site
is ``http://www.sdss.org/''.

The SDSS is managed by the Astrophysical Research Consortium for
the Participating Institutions. The Participating Institutions are The
University of Chicago, Fermilab, the Institute for Advanced Study, the
Japan Participation Group, The Johns Hopkins University, the Korean
Scientist Group, Los Alamos National Laboratory, the
Max-Planck-Institute for Astronomy, the Max-Planck-Institute
for Astrophysics, New Mexico State University, University of
Pittsburgh, University of Portsmouth, Princeton University, the United
States Naval Observatory, and the University of Washington.

\begin{table*}
\begin{center}
{Correlation Function Measurements}
\begin{tabular}{cccc}
\hline \hline
separation  & $\wpp(\rp)$ & $\xi(r)$ & Photometric \\
$(h^{-1}~\Mpc)$ & $(h^{-1}~\Mpc)$ & & Correction \\
\hline

$       0.010$ & $  7060(2300)$ & $ -63040(130000)$ & $    3.177$ \\
$       0.017$ & $ 20570(3500)$ & $471600(120000)$  & $   1.685$\\
$       0.026$ & $ 10950(2200)$ & $146300(49000)$   & $   1.331$\\
$       0.042$ & $  5387(850)$  & $35400(12000)$    & $  1.258$\\
$       0.066$ & $  3950(600)$  & $17100(4900)$     & $ 1.096$\\
$       0.105$ & $  2631(310)$  & $6865(1800)$      & $0.997$\\
$       0.166$ & $  1637(180)$  & $2366(590)$       & $1.000$\\
$       0.263$ & $  1161(78)$   & $1152(190)$       & $1.000$\\
$       0.417$ & $ 795.7(48)$   & $627.1(66)$       & $1.000$\\
$       0.660$ & $ 412.8(27)$   & $180.4(17)$       & $1.000$\\
$       1.047$ & $ 246.4(19)$   & $61.14(7.5)$      & $1.000$\\
$       1.659$ & $ 162.6(12)$   & $24.19(2.7)$      & $1.000$\\
$       2.629$ & $ 115.3(11)$   & $12.23(1.2)$      & $1.000$\\
$       4.167$ & $ 77.15(9.7)$  & $5.677(0.52)$     & $ 1.000$\\
$       6.604$ & $ 52.31(8.2)$  & $2.320(.19) $     & $ 1.000$ \\

\hline
\end{tabular}
\caption{\label{tab:data}
Measurements of the projected correlation function, $\wpp(\rp)$, and real-space correlation function, $\xi(r)$ for the LRG sample. The diagonal terms of the measurements error covariance matrices are given in parentheses.}
\end{center}

\end{table*}

\begin{table*}
\begin{center}

{ Properties of the Simulated Galaxies}
\begin{tabular}{ccccccc}
\hline \hline
$M_r$ & Half-Light Radius & $m_r$ & $u-r$ & $g-r$ & $r-i$ & $i-z$\\
\hline
$-22.5$ & $(10, 3.2)$ &	$17.867$ & $3.831$ & $1.722$ & $0.581$ & $0.955$\\
$-22.0$	& $(7,  2.24)$ & $18.367$ & $3.831$ & $1.722$ & $0.581$ & $0.955$\\
\hline

\end{tabular}
\caption{\label{tab:sim}Properties of the fake galaxies made for the photometry test. The half-light radius is given in comoving kpcs and arcseconds.}
\end{center}
\end{table*}

\end{document}